\documentclass[twocolumn,showpacs,preprintnumbers,amsmath,amssymb]{revtex4}

\usepackage{graphicx}
\usepackage{dcolumn}
\usepackage{bm}
\usepackage{CJK}

\begin{document}
\begin{CJK*}{GBK}{}

\title{Origin of negative differential thermal resistance in nonlinear systems}

\author{Dahai He$^{1}$} \email{dhhe@hkbu.edu.hk}
\author{Sahin Buyukdagli$^{1}$}
\author{Bambi Hu$^{1,2}$}
\affiliation{%
$^{1}$Department of Physics, Centre for Nonlinear Studies, and The
Beijing-Hong Kong-Singapore Joint Centre for Nonlinear and Complex
Systems (Hong Kong), Hong Kong Baptist
University, Kowloon Tong, Hong Kong, China\\
$^{2}$Department of Physics, University of Houston, Houston, Texas
77204-5005, USA
}%

\date{\today}

\begin{abstract}
Negative differential resistance in electronic conduction has been
extensively studied, but it is not the case for its thermal
counterpart, namely, negative differential thermal resistance
(NDTR). We present a classical Landauer formula in which the
nonlinearity is incorporated by the self-consistent phonon theory in
order to study the heat flux across a chain consisting of two weakly
coupled lattices. Two typical nonlinear models of hard and soft
on-site potentials are discussed respectively. It is shown that the
nonlinearity has strong impacts on the occurring of NDTR. As a
result, a transition from the absence to the presence of NDTR is
observed. The origin of NDTR consists in the competition between the
temperature difference, which acts as an external field, and the
temperature dependent thermal boundary conductance. Finally, the
onset of the transition is clearly illustrated for this model. Our
analytical calculation agrees reasonably well with numerical
simulations.
\end{abstract}

\maketitle
\end{CJK*}
\section{Introduction}
In the linear response regime, transport processes such as
transport of mass, momentum or energy can be described  by linear
laws of the form
\begin{equation}\label{Onsager}
j=\mu F,
\end{equation}
where $j$ and $F$ stand for the generalized flux and force,
respectively, and $\mu$ the transport coefficient. In other words,
the flux $j$ is a linearly increasing function of the external field
$F$, which is a well-known characteristic of Fick's law for mass
transport, Ohm's law for electron transport and Fourier's law for
heat transport. As the field $F$ becomes too strong, the system may
come into the nonlinear response regime, where the linear relation
\eqref{Onsager} is no longer valid since the transport coefficient
$\mu$ becomes itself field dependent. As a result, an interesting
phenomenon, i.e. negative differential resistance~(NDR) may take
place in a system in the strong-field regime where the flux
counter-intuitively decreases as the external field increases.

Since the pioneering observation in the tunnel diode by
Esaki~\cite{Esaki57}, NDR has been extensively studied for the
electronic transport, which led to widespread practical applications
in modern electronics~\cite{Sze81}. It is still an active research
topic to date, particularly at the atomic scale~\cite{Tu08,Pati08}.
However, its counterpart in the heat conduction problem, namely the
negative differential thermal resistance (NDTR) has been much less
studied. NDTR effect has been noticed in the studies of asymmetric
heat conduction (see Fig.~1 in Refs.~\cite{Li04}
and~\cite{Hu_SCPT06}), where it has been shown to be critical to
design a thermal diode with an enormous rectification factor. It has
also been shown that NDTR is crucial for a correct functioning of
lattice models of thermal transistor~\cite{Li06} and thermal logic
gates~\cite{Wang07}. It is already known that NDTR effect can be
qualitatively explained in terms of the overlapping of the
vibrational spectra of the interfacial particles~\cite{Li06,Wang07}.
On the other hand, is has been recently shown that in the model
presented in~\cite{Li06}, the NDTR effect will gradually disappear
as the system size increases or the properties of the interfacial
coupling change~\cite{Shao09,Zhong09}. For a clear understanding of
the mechanism underlying NDTR effect, it is thus imperative to
comprehend from a quantitative point of view the necessary
conditions for the occurring of NDTR effect.

As far as we know, a rigorous analytical approach to study heat
conduction in a non-integrable lattice system at the nonlinear
response regime has been so far unavailable. One usually has to rely
on numerical simulations. In the present study, we will develop a
phenomenological approach, in line with that in
Ref.~\cite{Hu_SCPT06}, to study heat flux through an ``interface"
between two weakly coupled anharmonic segments. We study two typical
models, which have hard and soft anharmonicity respectively. The
theoretical calculation based on the presented Landauer-like formula
yields results in reasonable agreement with the numerical
simulation. We will show that the intrinsic nonlinearity of the
system is necessary for the occurring of NDTR. It is further
illustrated that NDTR does not always occur in the presence of
nonlinearity. A transition from the absence to the presence of NDTR
with the increase of the nonlinearity is illustrated for
both models. A simple but physically appealing mechanism is proposed
in order to explain the origin of NDTR in the nonlinear systems. Our study of
NDTR provides indications of possible applications such as the
construction of thermal devices.

\section{Theoretical approach}
We study in this work the stationary heat current across a chain
consisting of two weakly coupled lattices,
\begin{equation} \label{Hamitonian}
H = H_{+}+\frac{K_{int}}{2} (x_{1}-x_0)^2+H_{-},
\end{equation}
where the Hamiltonian for the left and right segments are given by
\begin{equation}
H_+ = \sum^0_{i=-N/2+1} \frac {p_i^2}{2} +\frac {1}{2}
(x_{i+1}-x_{i})^2 +U_{+}(x_i)
\end{equation}
and
\begin{equation}
H_- = \sum^{N/2}_{i=1} \frac {p_i^2}{2} +\frac {1}{2}
(x_{i+1}-x_{i})^2 +U_{-}(x_i),
\end{equation}
respectively. $U_{\pm}(x)$ represent the onsite potential that will be specified below.
Two heat baths with temperatures $T_+$ and $T_-$ are connected to the
extremities of the left and right segment, respectively.  Note that
NDTR effect have been so far investigated only in spatially
asymmetric models~\cite{Li04,Hu_SCPT06,Li06,Wang07,Shao09,Zhong09}.
However, we will show in this study that NDTR can also take place in
a spatially symmetric model, i.e. for $U_{+}(x)=U_{-}(x)$.

In the case where the coupling $K_{int}$ between the left and right
segments is weak, the two segments will achieve two nearly
equilibrium states at temperatures $T_+$ and $T_-$, respectively.
Their vibrational motion can thus be approximately described
according to the self-consistent phonon theory
(SCPT)~\cite{Bruesch82,Dauxois93,Hu_SCPT06,He08a} with effective
harmonic Hamiltonians $H_{+}^{(0)}$ and $H_{-}^{(0)}$ of the form
\begin{equation}\label{HamitonianL}
H_{+}^{(0)} = \sum^0_{i=-N/2+1} \frac {p_i^2}{2} +\frac {1}{2}
(u_{i+1}-u_{i})^2 +\frac {f_{+}}{2}u_i^2,
\end{equation}
\begin{equation}\label{HamitonianR}
H_{-}^{(0)} = \sum^{N/2}_{i=1} \frac {p_i^2}{2} +\frac {1}{2}
(u_{i+1}-u_{i})^2 +\frac {f_{-}}{2}u_i^2,
\end{equation}
with $u_i=x_i-\langle x_i\rangle=x_i-\eta$. The effective
force constants $f_{\pm}$ are determined by numerically solving the
self-consistent equations
\begin{equation}\label{SCPG}
f_{\pm}=2\frac{\partial \langle U_{\pm}(x)
\rangle_{\pm}^{(0)}}{\partial \langle x^2 \rangle_{\pm}^{(0)} }
\end{equation}
Here $\langle...\rangle_{\pm}^{(0)}$ denotes the thermal
average with respect to the trial harmonic Hamiltonian
$H_{\pm}^{(0)}$ at the corresponding temperature $T_{\pm}$
and it is defined by
\begin{equation}\label{SCPthermalave}
\langle A(\mathbf{u}) \rangle_{\pm}^{(0)}
=\frac{\int{A(\mathbf{u})\exp{(-\beta_{\pm}
H_{\pm}^{(0)}(\mathbf{u}))}}\,\mathrm{d} \mathbf{u}}
{\int{\exp{(-\beta_{\pm}
H_{\pm}^{(0)}(\mathbf{u}))}}\,\mathrm{d}\mathbf{u}},
\end{equation}
for a given measurable $A(\mathbf{u})$, where
$\beta_{\pm}=(k_BT_{\pm})^{-1}$. Note that according to Eq.~\eqref{SCPthermalave},
$f_{\pm}$ is temperature dependent. The derivation of Eq.~\eqref{SCPG} can be found in the appendix of Ref.~\cite{He08a}. The renormalized normal-mode frequencies of phonons in each segment
can then be written as $\omega_{\pm}(q)=\sqrt{4\sin^2{(q/2)}+f_{\pm}}$, where $q$ stands for the
wave vector.

\begin{figure}
\includegraphics[width=1\linewidth]{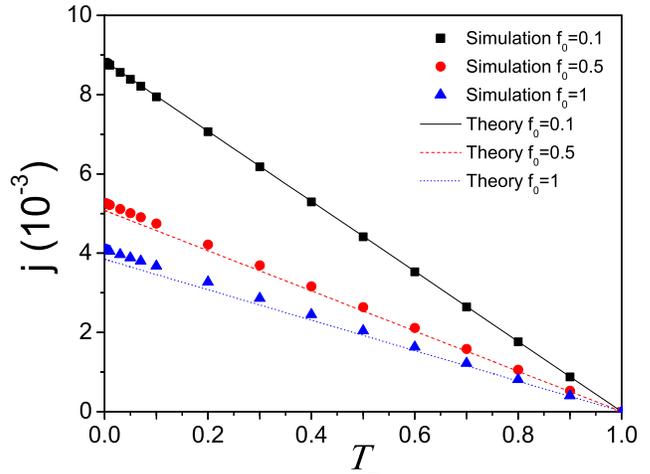}
\caption{\label{j_TR_PHI2}Heat flux $j$ as a function of $T_-$ for
the symmetric harmonic chain ($\lambda=0,f_{L0}=f_{R0}=f_0$). Here
$K_{int}=0.05$, $T_+=1$. The system size is $N=64$ for the simulation.
The linear dependence of $j$ on $T_-$ implies that the transmission
is temperature independent. }
\end{figure}
According to the Khalatnikov theory~\cite{Khalatnikov65}, the heat
flux can be regarded as the propagation of plane waves (phonons) with
various frequencies. Within this approach, the heat flux through the
system composed of the segments ~\eqref{HamitonianL}
and~\eqref{HamitonianR} can be determined by~\cite{He08}
\begin{equation}\label{LandauerC}
j =\frac{k_B(T_+-T_-)}{2\pi} \int_{\omega_{min}}^{\omega_{max}}
\mathcal{T}(\omega)\,\mathrm{d}\omega,\\
\end{equation}
where $\omega_{min}$ and $\omega_{max}$ correspond to the boundaries
of the overlap band of left and right phonon spectra, that is
$\omega_{min}=\max\left\{\sqrt{f_{+}},\sqrt{f_{-}}\right\}$ and
$\omega_{max}=\min\left\{\sqrt{4+f_{+}},\sqrt{4+f_{-}}\right\}$.
$\mathcal{T}(\omega)$ is the phonon transmission probability through
the interface. It is worth mentioning that Eq.~\eqref{LandauerC} is
similar in form to the celebrated Landauer formula
\begin{equation}\label{LandauerQ}
j =\frac{1}{2\pi} \int
\mathcal{T}(\omega)[\eta_{+}(\omega,T_+)-\eta_{-}(\omega,T_-)]\hbar\omega\,\mathrm{d}\omega,
\end{equation}
where $\eta_{\pm}=(e^{\beta_{\pm}\hbar\omega}-1)^{-1}$ are the
Bose-Einstein distribution functions. The Landauer
formula~\eqref{LandauerQ}, originated from the study of electron
transport~\cite{Landauer57}, describes the ballistic transport of
phonons in quantum systems~\cite{Rego98}. Considering the high
temperature limit (classical limit) where
$\eta_{\pm}\cong(\beta_{\pm}\hbar\omega)^{-1}$,
Eq.~\eqref{LandauerQ} reduces to Eq.~\eqref{LandauerC}. Note that
the quantum constant $\hbar$ is cancelled automatically in the
classical limit. Thus our equation ~\eqref{LandauerC} can be
considered as the classical form of the traditional Landauer
formula.

To find the transmission coefficient, we consider a plane wave
incident from the left, which is partly reflected by the interface
with amplitude $\bar{R}$ and partly transmits across the interface
with amplitude $\bar{T}$ into the right segment~\cite{Hu_SCPT06}.
The displacement of the $i$th particle from the equilibrium position
can be written as
\begin{equation}\label{wave}
u_i=\begin{cases}
(e^{Iki}+\bar{R} e^{-Iki})e^{-I\omega_{+}t}, & \text{if $i\leq0$}\\
\bar{T} e^{Iq(i-1)-I\omega_{-}t}, & \text{if $i\geq1$}
\end{cases}
\end{equation}
where $I$ is the imaginary unit. Thus the motion of the interface
particles can be described by the following equations
\begin{subequations}\label{motioneq}
\begin{align}
-\omega_{+}^2u_0&=u_{-1}+K_{int}u_1-(1+K_{int}+f_{+})u_0, \\
-\omega_{-}^2u_1&=u_{2}+K_{int}u_0-(1+K_{int}+f_{-})u_1.
\end{align}
\end{subequations}

If an acoustic matching condition $\omega_{+}=\omega_{-}=\omega$ is
satisfied, the solution of Eq.~\eqref{motioneq} gives the
transmission probability of Eq.~\eqref{LandauerC} in the form
\begin{equation}\label{transmission}
\begin{aligned}
&\mathcal{T}(\ \omega,f_{+}(T_+),f_{-}(T_-)\ ) =1-|\bar{R}^2| \\
&=\frac{C_2K_{int}^2}{C_1(1-2K_{int})+C_3K_{int}^2},
\end{aligned}
\end{equation}
where
\begin{subequations}
\begin{align}
C_1(\omega) &= (\omega^2-f_{+})(\omega^2-f_{-})\hspace{0.5mm},\\
C_2(\omega) &= \sqrt{C_1 (4+f_{+}-\omega^2)
(4+f_{-}-\omega^2)}\hspace{0.5mm}, \\
C_3(\omega) &= (C_1+C_2)/2 + 2\omega^2-f_{+}-f_{-}\hspace{0.5mm}.
\end{align}
\end{subequations}
Thermal transport is inhibited, i.e. $\mathcal{T}=0$ if the phonon
bands of the two segments are mismatched.
\begin{figure}
\includegraphics[width=.47\linewidth]{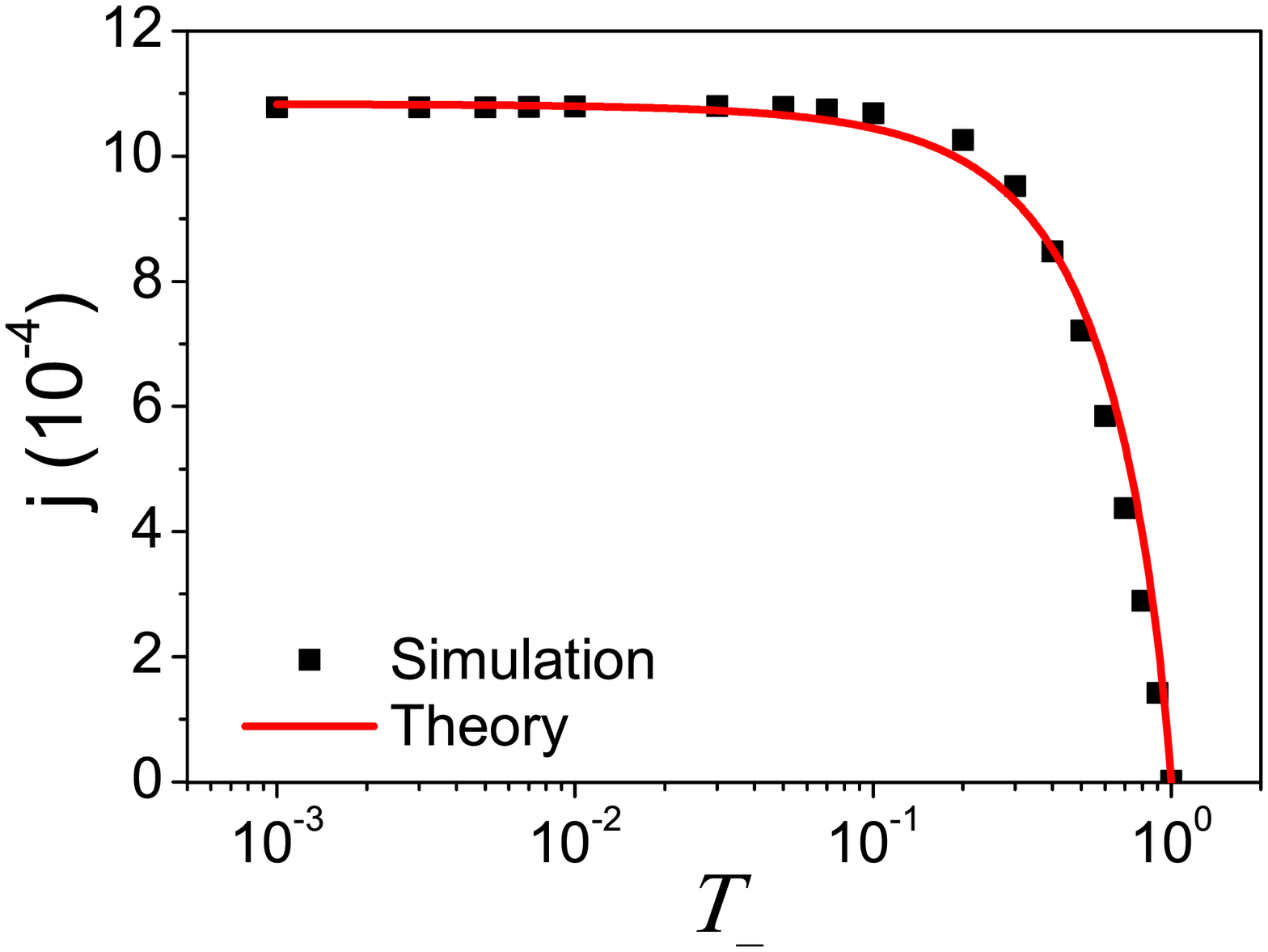}
\includegraphics[width=.47\linewidth]{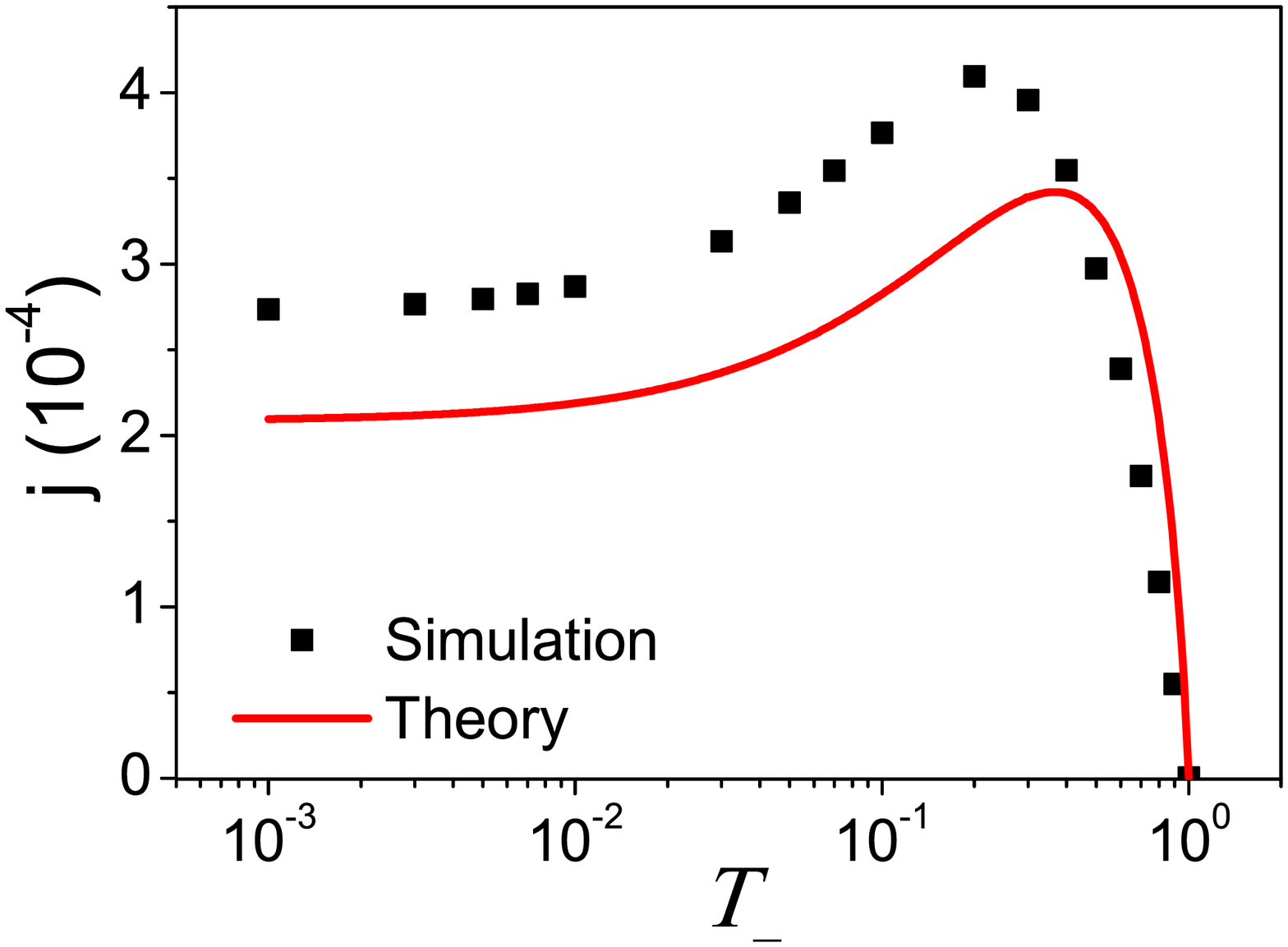}
\caption{\label{j_TR_lambdac}Heat flux $j$ as a function of $T_-$
for the $\phi^4$ model. Left: $\lambda=0.8$; Right: $\lambda=5$. In
both cases, $K_{int}=0.05$, $T_+=1$ and $N=64$ for the numerical
simulation.}
\end{figure}

In what follows, we will illustrate the results of the analytical
calculation based on Eq.~\eqref{LandauerC}. As a comparison,
non-equilibrium molecular dynamics (NEMD) simulations are performed
by applying Langevin heat baths at the two extremities of the chain. The
heat flux, whose definition can be found in Ref.~\cite{Lepri_rep},
is averaged over a long enough time so that
the system reaches the steady state regime, at which the local heat
flux is constant along the chain. During the simulations, $T_+$ was
fixed and the temperature difference $\Delta T=T_+-T_-$ was changed
by changing $T_-$.


\subsection{$\phi^4$ model} Let us first consider the $\phi^4$ model,
which is a typical bounding potential of ``hard'' anharmonicity ,

\begin{equation}\label{Potential_PHI4}
U_{\pm}(x)=\frac{f_{0}}{2}x^2+\frac{\lambda}{4} x^4.
\end{equation}
The effective force constants $f_{\pm}$ are determined by
numerically solving the self-consistent equations
\begin{equation}\label{SCP}
f_{\pm}=f_{0}+\frac{3\lambda k_BT_{\pm}}{\sqrt{f_{\pm}^2+4f_{\pm}}}.
\end{equation}
Note that $f_{\pm}$ regularly increases with increasing temperature.

Before discussing the nonlinearity effect, we will apply the above
analytical analysis to the harmonic model  $\lambda=0$. In this case,
the transmission~\eqref{transmission} is
temperature independent since $f_{\pm}=f_0$.  Fig.~\ref{j_TR_PHI2}
illustrates the heat flux $j$ as a function of $T_-$ for the harmonic
chain for several values of the harmonic constant $f_0$. By inspecting the
figure, we first notice that in the ballistic case, the simulation
agrees well with the analytical result that follows from the
classical Landauer formula~\eqref{LandauerC} ($k_B=1$). Furthermore,
it is seen that $j$ increases linearly when $T_-$ decreases, that is
when the temperature difference increases. As is expected, NDTR
cannot be observed in the harmonic model since there does not exist
any nonlinear response mechanism.

\begin{figure}
\includegraphics[width=0.9\linewidth]{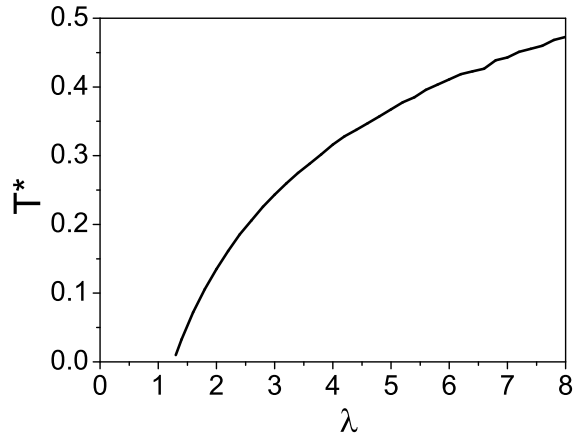}
\caption{\label{Tc}$\phi^4$ model: The turning point $T^*$, at which
the heat current exhibits a maximum, as a function of $\lambda$.
Nonzero $T^*$ indicates the presence of NDTR. The transition from
the absence to the presence of NDTR occurs at $\lambda_c\approx1$.}
\end{figure}

Furthermore,  Fig.~\ref{j_TR_lambdac} shows that if the
nonlinearity is present inside each segment ($\lambda\neq0$), the
formula~\eqref{LandauerC} still works reasonably well. By comparing
the left and right plots of Fig.~\ref{j_TR_lambdac}, it is seen
that a transition from the absence to the presence of NDTR occurs as
the nonlinearity $\lambda$ increases. We also present in
Fig.~\ref{Tc} the evolution of the turning point $T^*$ at which NDTR
effect manifests itself as a function of the nonlinearity. The on-set
of NDTR at $\lambda_c\approx1$ is clearly shown.

Note that Eqs.~\eqref{SCP} and~\eqref{LandauerC} implies the
following scaling relation
\begin{equation}\label{scaling}
j(T_{\pm},\lambda)=sj(T_{\pm}/s,s\lambda),
\end{equation}
where $s$ is an arbitrary scaling constant. The same scaling property
can be obtained from the equations of motion of the
model~\eqref{Hamitonian}(see \cite{Dhar_rep}). Eq.~\eqref{scaling}
indicates the equivalence between the temperature and the
nonlinearity. Thus a similar transition for fixed $\lambda$ can be
observed as $T_+$ increases, which will be verified both from numerical simulations
and our analytical approach. In fact, NDTR takes place if $\lambda>\lambda_c$ (or
$T_+>T_c$).


\subsection{On-site Morse model} Now we consider a model which consists
of two weakly coupled symmetric nonlinear lattices with an on-site
Morse potential given by
\begin{equation}\label{Potential_Morse}
U_{\pm}(x)=D[\,\exp{(-\alpha x)}-1]^2.
\end{equation}
Model~\eqref{Potential_Morse}  was introduced in order to investigate the
DNA denaturation process~\cite{Peyrard89}. The anharmonicity of the
model is ``soft" since the Morse potential is bounded for
$x\rightarrow \infty$.

\begin{figure}
\includegraphics[width=0.9\linewidth]{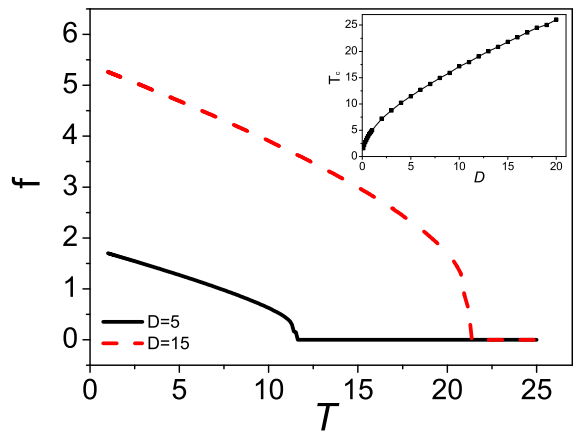}
\caption{\label{fT}Effective force constant $f$ as a function of
the temperature for the model~\eqref{Potential_Morse}. Inset: The
critical temperature $T_c$, above which $f$ vanishes, against the
nonlinearity $D$. }
\end{figure}

The effective force constants $f_{\pm}$ are obtained by numerically
solving the following self-consistent equations
\begin{equation}\label{SCP_Morse}
f_{\pm}=2\alpha^2D\exp{\left(-\frac{\alpha^2
k_BT_{\pm}}{\sqrt{f_{\pm}^2+4f_{\pm}}}\right )}.
\end{equation}
It should be noted that $f_{\pm}$ decrease as the temperature increases
for a soft anharmonic potential like Eq.~\eqref{Potential_Morse}, as
shown in Fig.~\ref{fT}. This is different from that of a model with a
hard anharmonicity like Eq.~\eqref{Potential_PHI4}, for which
$f_{\pm}$ monotonically increases as the temperature increases. One can
see from Fig.~\ref{fT} that there exists a critical temperature
$T_c$, above which the force constant $f$ vanishes. It means that
the on-site potential can be neglected once the thermal energy of
the particles overcomes the potential energy and then the system
behaves like a harmonic chain. The inset of Fig.~\ref{fT}
shows that the critical temperature increases with the nonlinearity of
the system, which reflects the fact that the deeper the
potential well is, the larger is the thermal energy needed to
overcome the potentiel barrier.

\begin{figure}
\includegraphics[width=.47\linewidth]{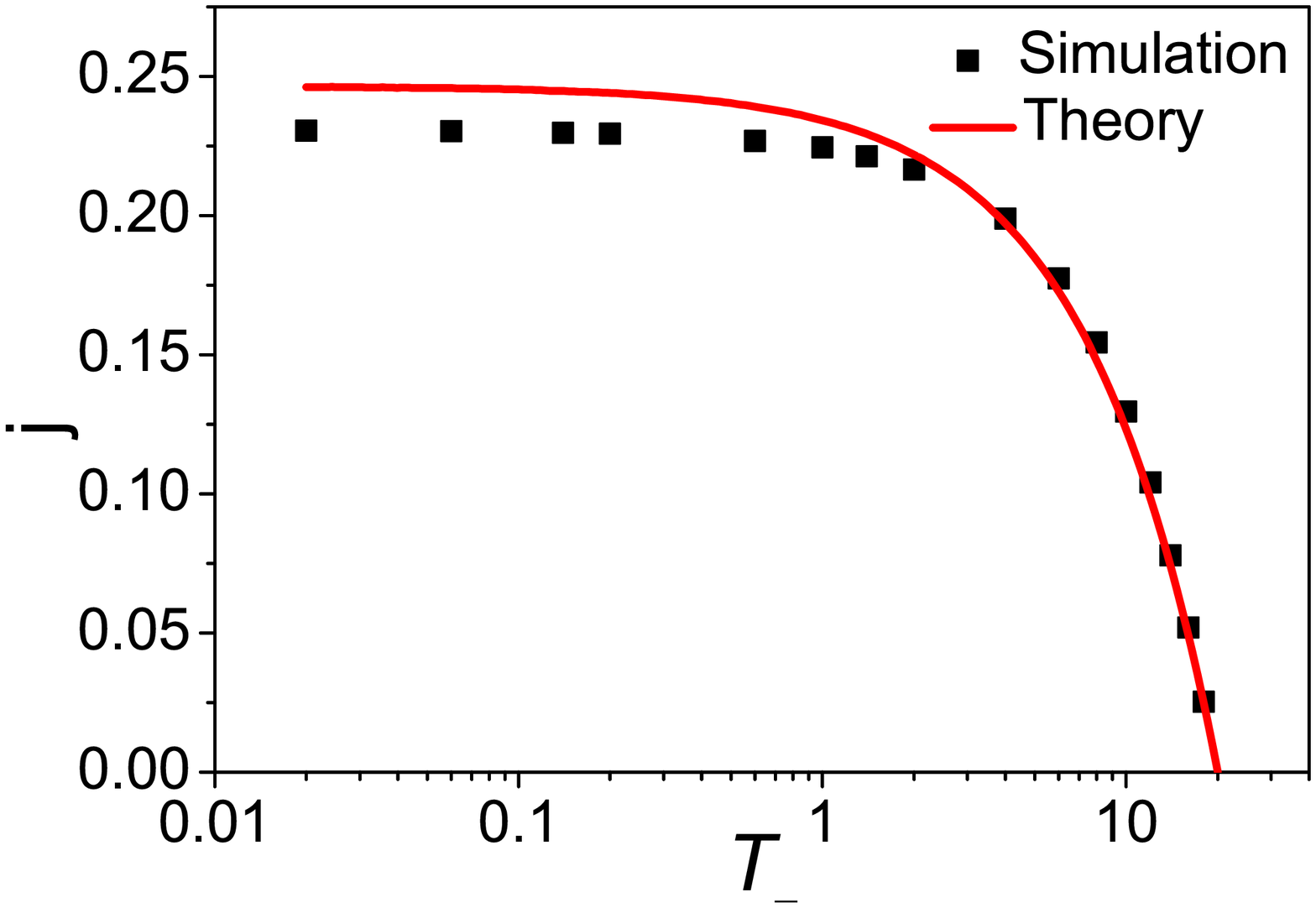}
\includegraphics[width=.47\linewidth]{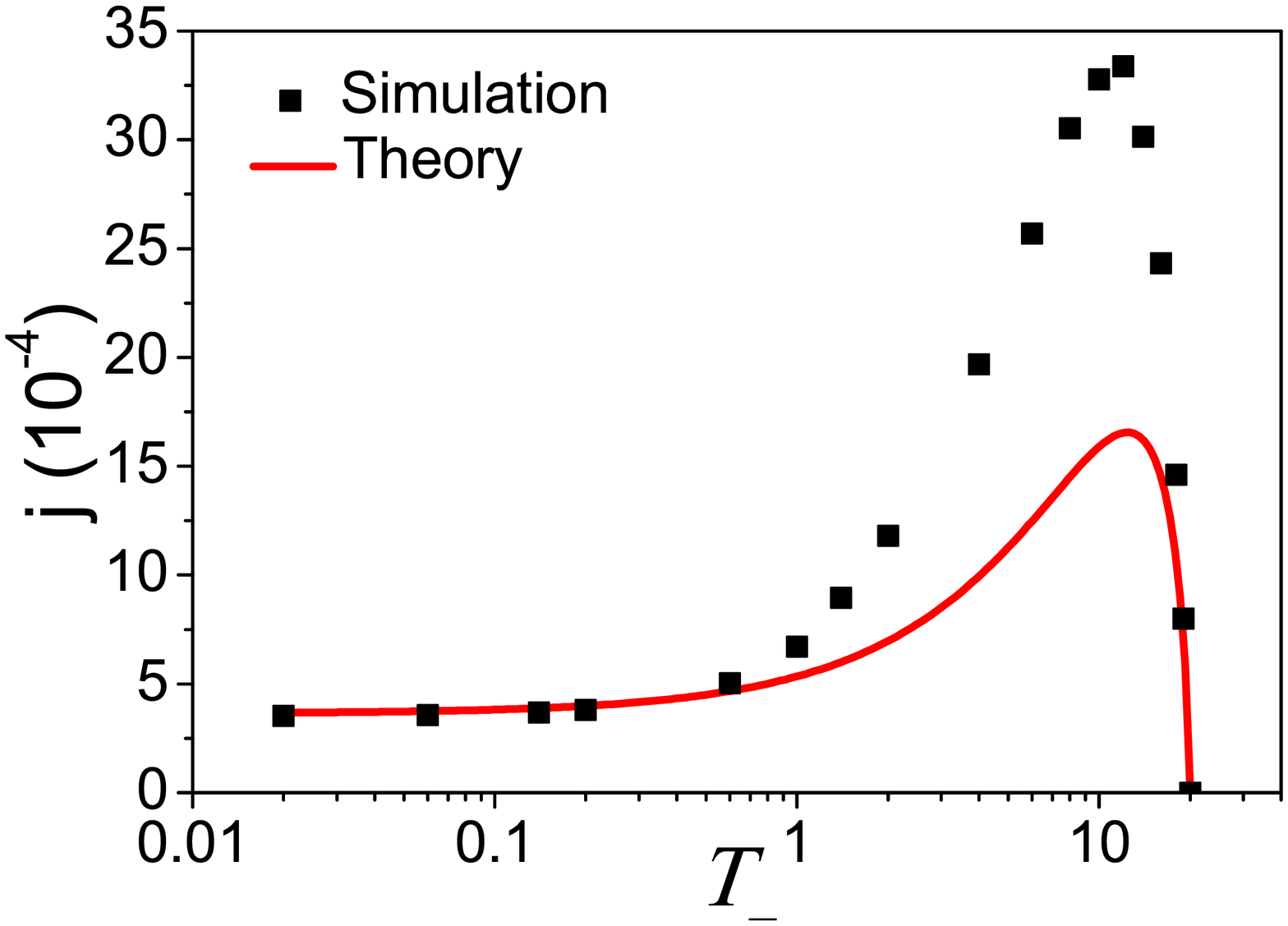}
\caption{\label{Morse_j_TR}Heat flux $j$ as a function of $T_-$ for
the Morse model. Left: $D=5\times10^{-5}$; Right: $D=15$. In both
cases, $\alpha=0.426$, $K_{int}=0.05$, $T_+=20$ and $N=64$ for the
numerical simulation.}
\end{figure}

We then use Eq.~\eqref{LandauerC} to compute the heat flux and
compare the analytical result with numerical simulations, as shown in
Fig.~\ref{Morse_j_TR}. One can see that since the nonlinearity
$D$ is weak, the heat flux increases
monotonically with increasing $\Delta T$ . Nevertheless, NDTR occurs as $D$ becomes large enough.
Like $\phi^4$ model, Fig.~\ref{Morse_j_TR} indicates that as the nonlinearity
increases, the soft potential model also exhibits a transition from positive differential thermal
resistance (PDTR) to NDTR.

\begin{figure}
\includegraphics[width=0.9\linewidth]{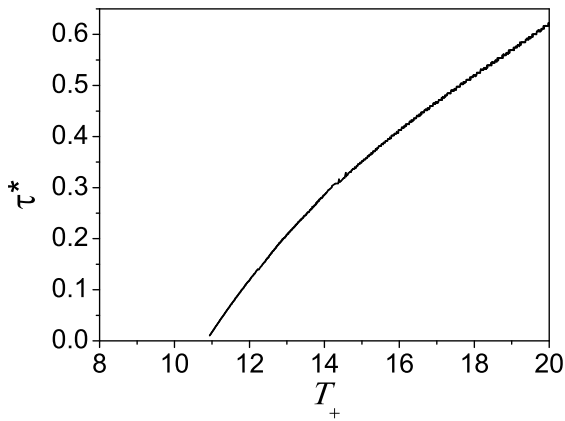}
\caption{\label{Tc_morse}Morse model: The scaled turning temperature
$T^*/T_+$, at which the heat current exhibits a maximum, as a function
of $T_+$. Here $D=15$,$\alpha=0.426$.}
\end{figure}
A simple scaling relation like Eq.~\eqref{scaling} for the
nonlinearity and the temperature is inexistent for the Morse model.
However, increasing the nonlinearity is qualitatively equivalent to
increasing the temperature. Thus a similar transition from PDTR to
NDTR as the temperature increases can be expected. In
Fig.~\ref{Tc_morse}, we calculate the scaled turning point
$\tau^*\equiv T^*/T_+$, at which the heat flux exhibits a maximum.
Non-zero $\tau^*$ indicates the presence of NDTR. The analytical
result shows the on-set of NDTR at $T_+\approx11$ for the
dissociation threshold fixed at $D=15$.

However, one should note that the SCPT fails in the vicinity of the
melting transition for the present soft potential model, as pointed out in
Ref.~\cite{Dauxois93}. The reason for the failure of the variational approach
is the fact that the half-bounded
potential~\eqref{Potential_Morse} cannot be simply approximated by
the trial harmonic potential with a temperature-dependent force
constant. This pecularity prevents us from modelling the crucial role of the nonlinearity in this
regime using the SCPT. This point will be discussed
in detail in the next section. Although it is clear that the quantitative
agreement with the simulation result is poor, we emphasize that
the present approach goes far beyond the traditional Landauer
approach in its ability to characterize the nonlinear response
regime, for which a transition from PDTR to NDTR is illustrated at
least in a qualitative manner.

\section{Physical Mechanism}
\begin{figure}
\includegraphics[width=0.47\linewidth]{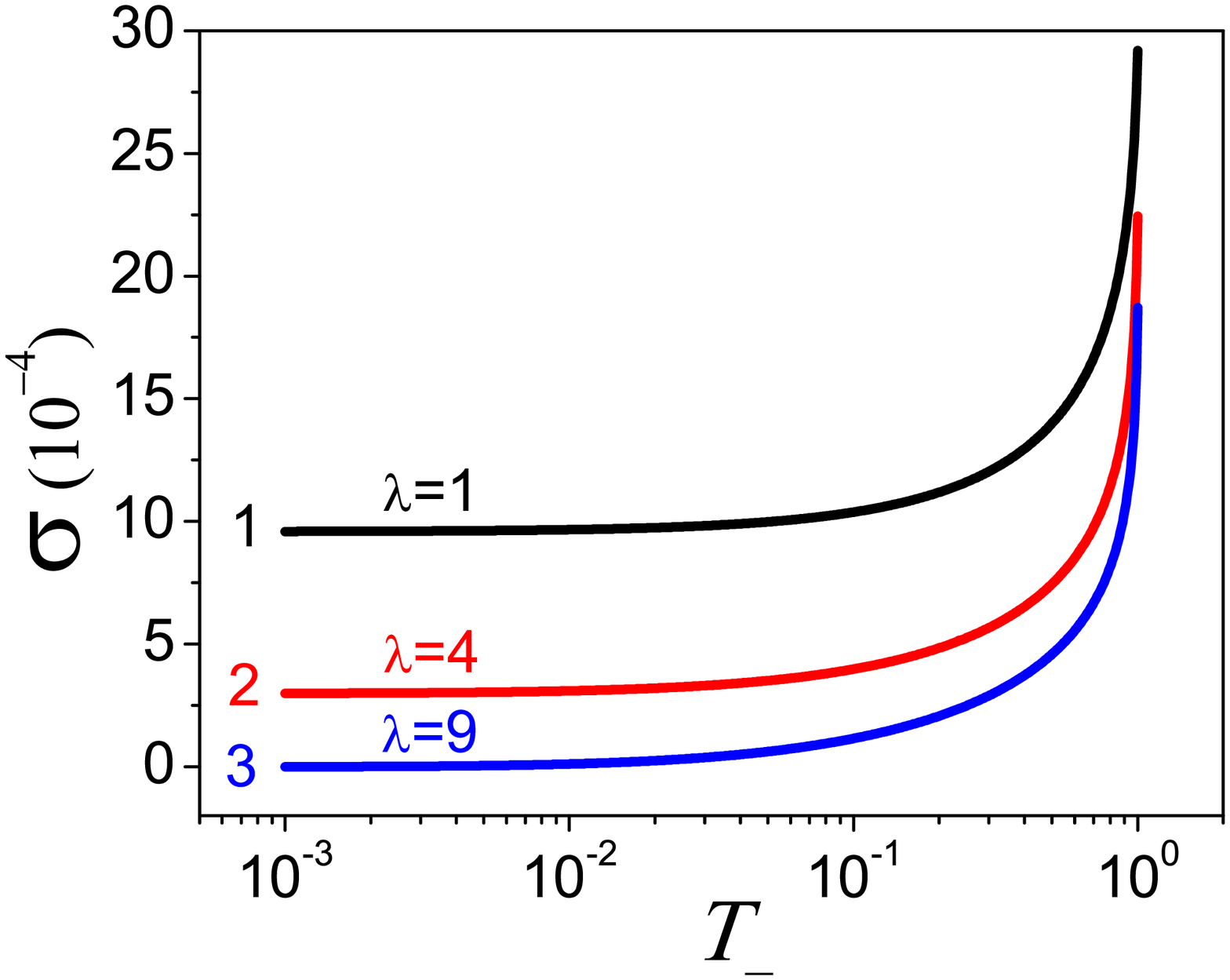}
\includegraphics[width=0.47\linewidth]{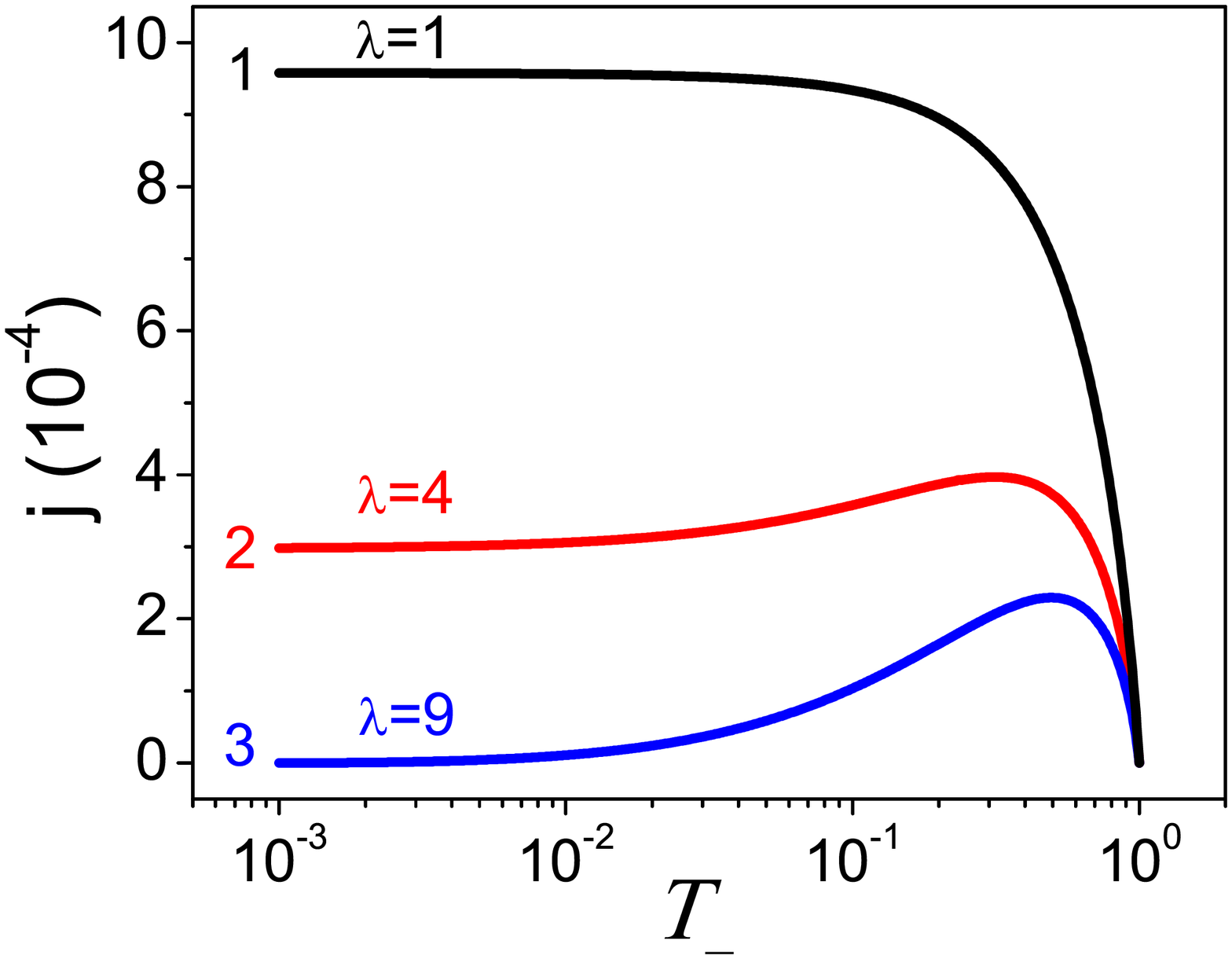}
\caption{\label{j_TR_lambda}$\phi^4$ model: The thermal boundary
conductance $\sigma$ (left plot) and the heat flux $j$ (right plot)
against $T_-$ for different values of $\lambda$. Lines 1, 2, 3
correspond to $\lambda=1$, 4, 9, respectively. One can see that
nonlinearity $\lambda$ has an important effect on the behavior of
the heat flux due to the interchange of the dominant role in
Eq.~\eqref{ohmic}. Here $T_+=1$, $K_{int}=0.05$.}
\end{figure}

\begin{figure}
\includegraphics[width=0.47\linewidth]{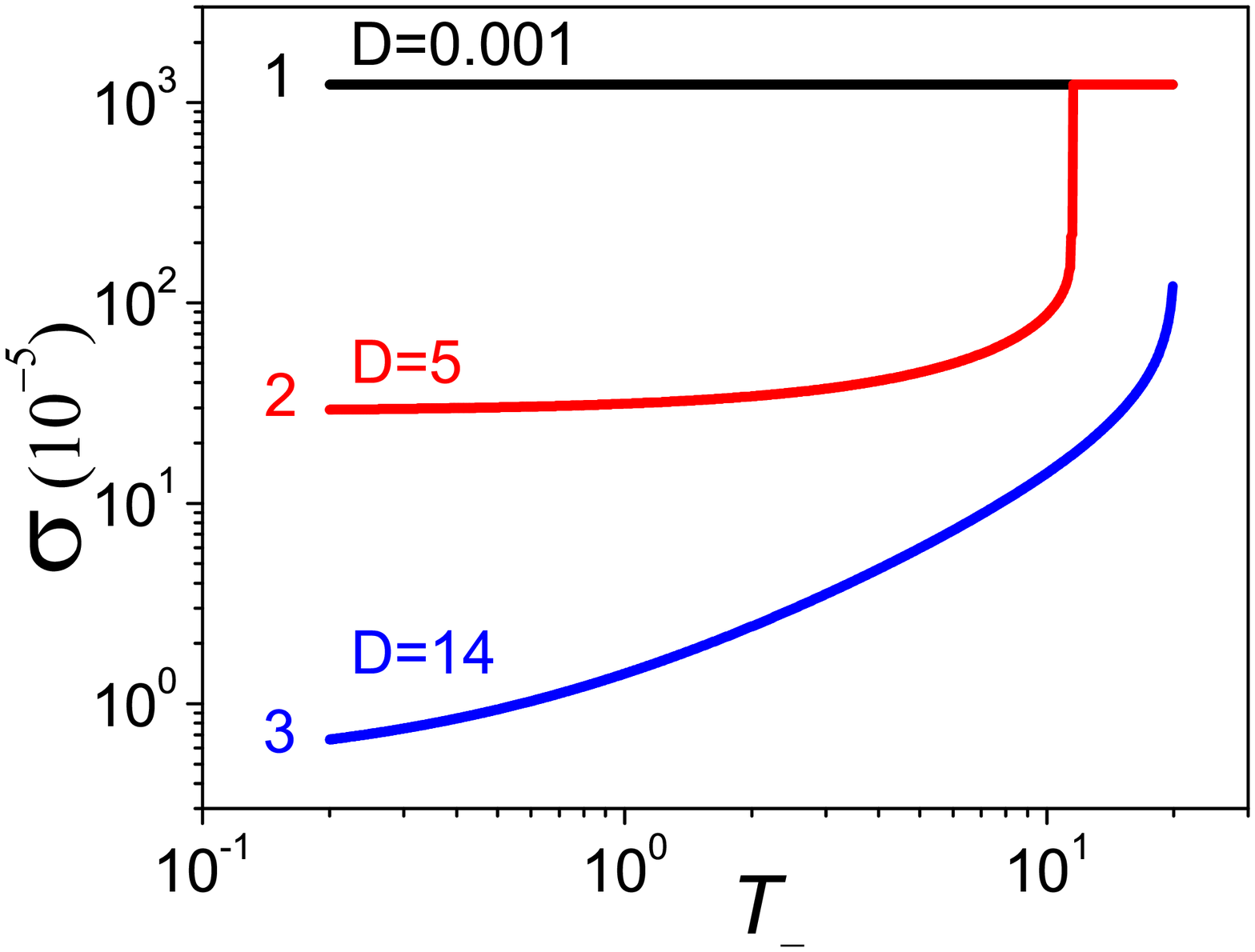}
\includegraphics[width=0.47\linewidth]{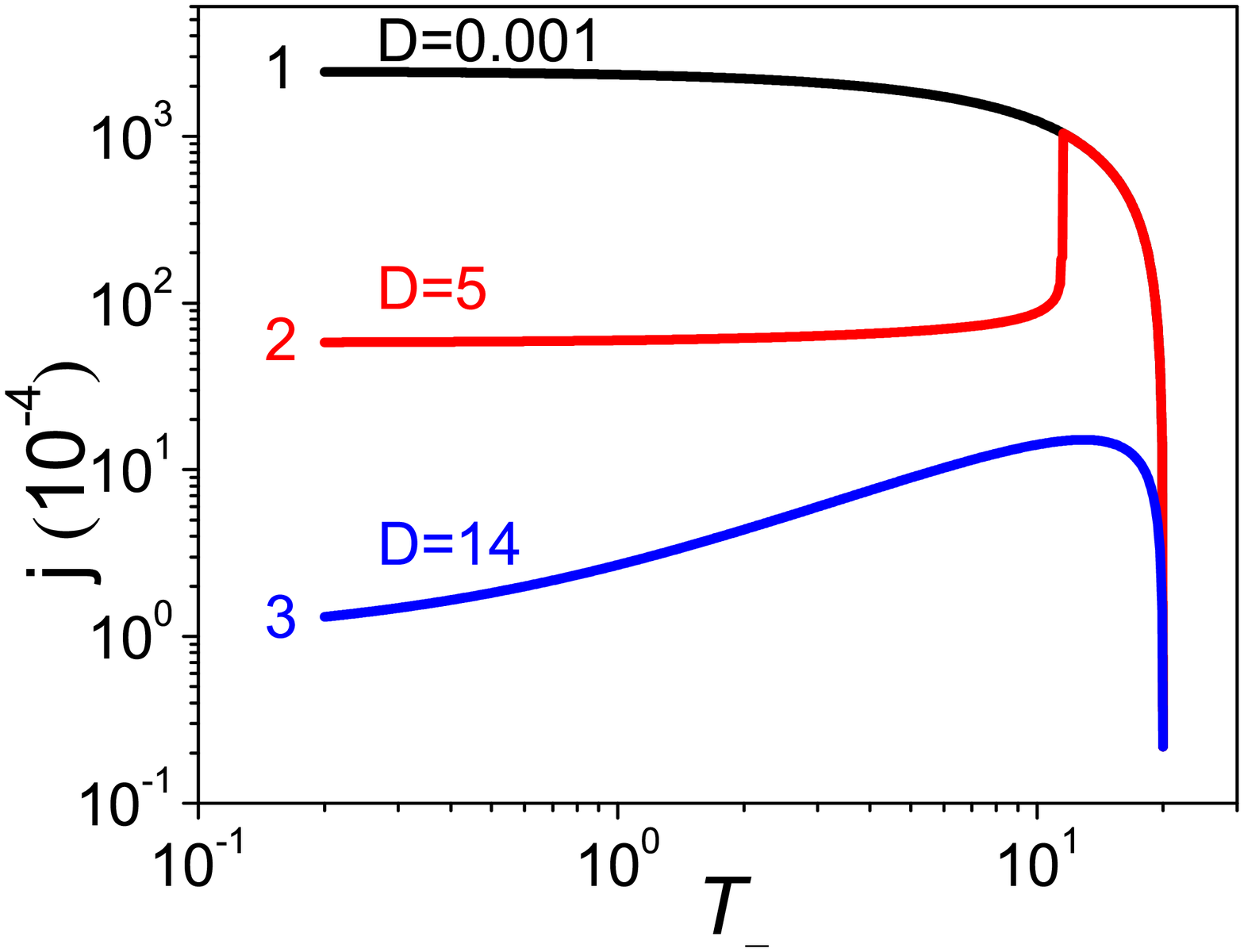}
\caption{\label{j_TR_lambda_morse} Morse model: The thermal boundary
conductance $\sigma$ (left) and the heat flux $j$ (right) as a
function of $T_-$ for different values of $D$. Lines 1, 2, 3
correspond to $D=0.001, 5, 18$, respectively. Here $T_+=20$,
$K_{int}=0.05$.}
\end{figure}

The results presented so far give rise to the following question:
what is the origin of NDTR in the above models? We will show that the
classical Landauer equation~\eqref{LandauerC} yields a simple and
intuitive explanation. One should note that the temperature
discontinuity at the virtual interface(the site $x_0$) indicates the
existence of the thermal boundary resistance (or conductance, see
Ref.~\cite{Swartz89}), and it plays a crucial role for the heat
conduction in our weakly-coupled model. Defining the effective
thermal boundary conductance by
\begin{equation}\label{TBC}
\sigma =\frac{k_B}{2\pi} \int_{\omega_{min}}^{\omega_{max}}
\mathcal{T}(\omega) \,\mathrm{d}\omega,
\end{equation}
Eq.~\eqref{LandauerC} can be rewritten as
\begin{equation}\label{ohmic}
j=\Delta T\sigma,
\end{equation}
which is similar in form to Ohm's law for electrical conduction. The
simple relation ~\eqref{ohmic} suggests that there exists mainly two
contributions to the temperature dependence of the heat current for
a two-segment system. The first contribution comes from the
temperature difference $\Delta T$, which acts as an external thermal
force and yields the regularly increasing behavior of $j$ with
decreasing $T_-$. The second contribution is due to the thermal
boundary conductance $\sigma$. One can see from
Fig.~\ref{j_TR_lambda} that unlike $\Delta T$, $\sigma$ is an
increasing function of $T_-$. The widening of the overlap band of
the vibrational spectrum of segments $L$ and $R$, or
$\Delta\omega=\omega_{max}-\omega_{min}$, is mainly responsible for
this increasing behavior of the thermal conductance. Consequently,
the origin of NDTR effect is basically the competition between the
growing ``external field" $\Delta T$ and the diminishing overlap
band $\Delta\omega$ as $T_-$ decreases. NDTR thus occurs below the
temperature $T^*$ at which the opposite behavior of both
contributions exactly compensate each other and it takes place if
and only if $\sigma$ becomes dominant for $T_-<T_+$.
Fig.~\ref{j_TR_lambda} shows the apparition of NDTR effect as one
considers different nonlinearity parameters $\lambda=1, 4, 9$ for
the $\phi^4$ model. One can note that for segments with $\lambda$
large enough, $j$ vanishes as $T_-$ decreases due to the mismatch of
the phonon bands. We also plot in Fig.~\ref{j_TR_lambda_morse} the
thermal boundary conductance $\sigma$ and the corresponding heat
flux $j$ for the Morse model, which displays a similar behavior. We can thus conclude
that the proposed mechanism for the occurring of NDTR is valid
for both hard and soft models. For the harmonic
system, $\sigma$ is exactly temperature independent, leading to the
linear behavior observed in Fig.~\ref{j_TR_PHI2}.

One should note that the curve 2 of Fig.~\ref{j_TR_lambda_morse}
exhibits  a jump at $T_-=T_c$ . For $T_-\geq T_c$, the effective
force constant $f$ vanishes and the phonon frequency becomes
temperature independent. The system thus behaves like a harmonic
chain, characterized by a temperature independent thermal boundary
conductance $\sigma$ and a linear behavior of the heat flux. The
occurrence of the jump and the exact linear behavior of the heat
flux as $T_-\geq T_c$ are inconsistent with the numerical
simulation. As discussed in the last section, this artificial result
lies in the incapability of the SCPT to model the transition at the
vicinity of $T_c$ as shown in Fig.~\ref{fT}. Since $T_c$ increases
with $D$, the artificial jump disappears provided $T_+<T_c(D)$,
which is illustrated in the right plot of Fig.~\ref{Morse_j_TR}.

\section{SUMMARY}
In summary, we presented a classical Landauer formula to study NDTR effect
in typical lattice models. It was shown that NDTR cannot occur
in a harmonic lattice, for which the linear relation~\eqref{Onsager}
is generally valid no matter how large the temperature difference
is. In the presence of anharmonicity, one can observe a transition
from the absence to the presence of NDTR as the nonlinearity is
increased for both hard and soft potentials. The NDTR effect is
basically due to the competition between the increasing behavior of
the external field and the decreasing behavior of the effective
thermal boundary conductance of the chain. The transition in
Figs.~\ref{Tc} and~\ref{Tc_morse} indicates that NDTR may be
controlled by adjusting the parameters of the system or the
temperature of heat baths, whose utility for nanoscale applications
is clear.

It is imperative to clarify why we are allowed to apply such a
seemingly ballistic transport formula~\eqref{LandauerC} to calculate
the heat flux  through a nonlinear system in the strong field
regime. For the particular model presented, even though the whole
system is at strong external field, each segment is still close to
its corresponding equilibrium state thanks to the weak coupling, and
can thus be approximately described by SCPT. It should be stressed
that the analytical estimation, which is based on the local thermal
equilibrium of the segments, holds only if the coupling $K_{int}$ is
weak enough. For strong coupling, both segments get far from thermal
equilibrium and SCPT can not be applied anymore to deal with the
nonlinearity.

Finally, it is worth giving a comment about the relation between
asymmetric heat conduction and NDTR. Note the following two facts:
1) Asymmetric heat conduction results from the intrinsic spatial
asymmetry of the model, which is not necessary for the occurring of
NDTR as shown in this study; 2) One can observe asymmetric heat
conduction without the occurring of NDTR as long as the applied
temperature difference is moderate. It seems that the NDTR, as a
field-induced effect, is neither a sufficient nor a necessary
condition for thermal rectification.

\begin{acknowledgments}
We acknowledge the helpful discussions with members of Centre for
Nonlinear Studies at Hong Kong Baptist University.
\end{acknowledgments}

\end{document}